\documentclass[a4paper,fleqn,usenatbib]{mnras}
\usepackage[T1]{fontenc}
\usepackage{ae,aecompl}

\usepackage{graphicx}	
\usepackage{amsmath}	
\usepackage{amssymb}	
\usepackage{bm}
\usepackage{subfig}
\usepackage{times}
\usepackage{color}
\usepackage{soul}

\newcommand{\bhac}{\texttt{BHAC}~}

\newcommand{\eg}{e.g.,~}
\newcommand{\ie}{i.e.,~}

\title[Magnetic reconnection around black holes]{Magnetic reconnection
  and plasmoid formation in three-dimensional accretion flows around
  black holes}

\author[A. Nathanail et al.]{Antonios Nathanail$^{1,4}$\thanks{E-mail:
    antonionitoni@hotmail.com,anathanail@phys.uoa.gr}, Vasilis Mpisketzis$^{1}$, Oliver
  Porth$^{2}$, Christian M. Fromm$^{3,4,5}$,  \newauthor  and Luciano Rezzolla$^{4,6,7}$ \\
  $^{1}$Department of Physics, National and Kapodistrian University of
  Athens, Panepistimiopolis, GR 15783 Zografos, Greece \\
  $^{2}$Astronomical Institute Anton Pannekoek, Universeit van
  Amsterdam, Science Park 904, 1098 XH, Amsterdam, The Netherlands \\
  $^{3}$ Institut f\"ur Theoretische Physik und Astrophysik,
  Universit\"at W\"urzburg, Emil-Fischer-Strasse 31, 97074 W\"urzburg,
  Germany \\
  $^{4}$Institut f\"ur Theoretische Physik, Goethe Universit\"at
  Frankfurt, Max-von-Laue-Str.1, 60438 Frankfurt am Main, Germany \\
  $^{5}$Max-Planck-Institut f\"ur Radioastronomie, Auf dem H\"ugel 69,
  D-53121 Bonn, Germany \\
  $^{6}$School of Mathematics, Trinity College, Dublin 2, Ireland\\
  $^{7}$Frankfurt Institute for Advanced Studies, Ruth-Moufang-Str. 1,
  60438 Frankfurt am Main, Germany}

\usepackage{hyperref}
\hypersetup{
  citecolor=cyan,      
  filecolor=magenta,      
  urlcolor=magenta            
}

\begin{document}
\label{firstpage}
\pagerange{\pageref{firstpage}--\pageref{lastpage}}
\maketitle

\begin{abstract}
Magnetic reconnection is thought to be one of the main energy-dissipation
mechanisms fueling energy to the plasma in the vicinity of a black
hole. Indeed, plasmoids formed through magnetic reconnection may play a
key role in $\gamma$-ray, X-ray and near-infrared flares from the black
hole at the center of our galaxy, SgrA*. We report the results of
three-dimensional general-relativistic ideal and resistive
magnetohydrodynamics simulations modelling magnetic reconnection in
accretion flows around astrophysical black holes. As an important
difference with similar works, our accretion discs have an initial
dipolar magnetic-field configuration with loops of alternating
polarity. We show that current sheets are formed and destroyed rapidly in
the turbulent environment of black-hole accretion. Plasmoids are 
	formed from current sheets close to the event
horizon, in a region of $\sim2-15$ gravitational radii. We further
quantify the magnetic dissipation and the process of energy transfer to
the plasmoids, reporting the reconnection rate, the relative current
density with respect to the local magnetic field, and the size of the
plasmoids. We find that plasmoids gain energy through reconnection and
heat up to relativistic temperatures, with the largest ones being
sufficiently energetic to leave the black hole near the polar
regions. During their evolution, plasmoids are stretched and elongated,
becoming disrupted when the shear is sufficiently large, although some
plasmoids survive as well-distinguished structures at distances of
$\sim30-40$ gravitational radii from the black hole. Finally, we find
that in some cases the plasmoids acquire a super-Keplerian azimuthal
velocity, as suggested by recent observations of flares from
Sgr~A*.  \end{abstract}

%

\section{Introduction} 
\label{sec:intro}
%

The supermassive black hole in the Galactic center, Sgr~A*, is the
closest of its kind and is the subject of several campaigns of
multi-wavelength observations since it is an excellent laboratory for
accretion physics in the extreme-gravity regime. Over the years,
unparalleled insight on the accretion flow in the immediate environment
of the central black hole has been obtained
\citep{Falcke1998,Doeleman2008, Dodds-Eden2009, Reid2009, Johnson2015b,
  Broderick2016}. On a daily basis bright flares in the X-rays and in the
near-infrared (NIR) are observed \citep{Baganoff2001, Genzel2003,
  Marrone2007, Witzel2018}.
It has been suggested that flares may be the result of magnetic
reconnection in the vicinity of the central black hole
\citep{Markoff2005, Broderick2006, Yuan2009, Barriere2014,
  Haggard2019}. Flares are observationally associated with an increase of
almost two orders of magnitude in the typical NIR emission
\citep{Ponti2017,Do2019}, but such an increase is difficult to be
accounted for in terms of a significant change in the accretion flow
\citep{Ressler2018}. On the other hand, orbital motion originated at the
inner edge of the accretion flow, around the supermassive black hole,
Sgr~A*, was recently detected from the \citet{Abuter2018b}, thus
leading support to the idea that localised emission does take place in
the accretion flow.

Promising phenomenological models that match the overall observables of
Sgr~A* \citep[see, \eg][]{Yuan2003} lack the detailed description of the
accretion flow and must be enhanced with theoretical description of the
flaring activity observed in Sgr~A*. The latter can be accomplished by
incorporating magnetic reconnection as a channel to dissipate the
magnetic energy and naturally give rise to localised emission, modeled as
hot-spots or plasmoids, and ultimately leading to the observed flares
\citep{Broderick2005, Dodds-Eden2010, Younsi2015, Ball2016, Li2017flare,
  Gutierrez2019, Baubock2020}.

Global simulations can accurately describe the accretion dynamics but
lack the detailed microphysical description of particle acceleration in
the current sheets that are formed and abruptly destroyed close to the
event horizon. Large current sheets of alternating magnetic-field
polarity have been found to be very efficient in particle acceleration.
More specifically, the initial magnetic field tends to fragment at the
location of the current sheet and subsequently produces chains of
magnetic ``plasmoids'' or ``magnetic islands'' \citep{Loureiro2007,
  Uzdensky2010, Fermo2010, Huang2012, Loureiro2012, Takamoto2013}. We
recall that plasmoids are magnetic-flux bundles, which, in the absence of
a bulk motion, may become quasi-spherical blobs due to a fast tearing
instability; these blobs then contain and trap charged particles that can
be accelerated to relativistic energies. Overall, plasmoids are
characterised by possessing a rather large magnetisation
$\sigma:=B^2/\rho$, that is, a large ratio between the magnetic --
$B:=\sqrt{B^iB_i}$ is the magnetic-field strength, with $B^i$ the
magnetic-field components -- and the rest-mass energies. A
phenomenological cutoff of $\sigma \gtrsim 0.3$ was found in
two-dimensional (2D) general-relativistic magnetohydrodynamics (GRMHD) in
the ideal limit \citep{Nathanail2020, Ripperda2020}, a result that is
consistent with local, particle-in-cell (PIC) simulations
\citep{Sironi2016, Li2017PIC, Kagan2018,Petropoulou2018,
  Petropoulou2019}. Indeed, although restricted to microscopical scales,
PIC simulations have provided a significant insight in the process of
magnetic reconnection in current sheets, where plasmoids and plasmoid
chains are naturally accelerated and produce distributions of non-thermal
electrons \citep{Sironi2014, Guo2014PIC, Guo2015, Werner2016, Rowan2017,
  Ball2018a, Rowan2019, Guo2021, Guo2021b}.

The global picture of the reconnection processes taking place in the
turbulent accretion environments near a black hole can be drawn only by
exploiting accurate and long-term GRMHD simulations \citep{Chan2015,
Dexter2009b, Dodds-Eden2010, Ball2016, Dexter2020, Porth2021,
Ripperda2021}. Such simulations have recently provided a significant
boost to the study and understanding of the occurrence and on the impact
that magnetic reconnection has in accretion flows \citep{Ball2018,
Qian2018, Kadowaki2018, Vourellis2019, White2020, Cemeljic2020,
Dihingia2021, Chaskina2021, Scepi2021}. In particular, the production of energetic
plasmoids that tend to orbit in the vicinity of the black hole were
observed in high-resolution GRMHD simulations both in the ideal limit
\citep{Nathanail2020} -- in which case the resistivity is purely of
numerical origin -- and in general-relativistic resistive (GRRMHD)
simulations -- where the resistivity is instead physical
\citep{Ripperda2020}.

These studies confirmed the expectation that plasmoids are generated
close to the event horizon from magnetic reconnection events. The
plasmoids can have a size of a few gravitational radii and the
``reconnection rate'' -- which is defined in terms of the ratio between
the velocity of the plasma inflowing into the current sheet and that
leaving the sheet -- was found to be between $0.01$ and $0.03$
\citep{Ripperda2020} confirming previous relativistic studies, both
numerical and analytic \citep{DelZanna2016, Ripperda2019b,
  Bhattacharjee2009, Uzdensky2010}. The production of plasmoids in the
simulation does not alter or enhance the variability of the accretion
flow, but it may be responsible for observed variability (or even
flaring) in the NIR or the X-rays \citep{Nathanail2020}. After they are
formed, plasmoids can steadily acquire additional energy through
continued magnetic reconnection, as they evolve and cool down. Moreover,
in an accretion flow originating from a disc, plasmoids have in general a
non-zero angular momentum and hence can potentially be related to
observations of orbiting hot spots near the galactic center
\citep{Abuter2018b}.

In our previous work in which we have analysed the production and
evolution of plasmoids in accretion flows onto black holes
\citep{Nathanail2020}, the restriction to a 2D setup and thus the
assumption of axisymmetry has prevented us from a direct comparison with
the observations. Furthermore, the simulations by \citet{Nathanail2020}
could not capture the development of non-axisymmetric instabilities that
affect both the production, but, more importantly, the evolution of the
plasmoids. The most important aspect of this evolution is that plasmoids,
as they move outwards from the vicinity of the black hole, are stretched
and elongated, sometimes even being disrupted when the shear is
sufficiently large.

We here present the results of three-dimensional (3D) GRMHD simulations
of magnetic reconnection in the vicinity of a black hole. More
specifically, discuss and compare the same simulation setup when explored
solving either the equations of GRMHD either in the ideal MHD limit or in
the presence of a finite and physical resistivity. In both cases, we find
that the production of plasmoids and of plasmoid chains is comparable and
similar to one obtained in the 2D simulations of \citet{Nathanail2020}.
On the other hand, the 3D simulations also reveal important differences
in the evolution of the plasmoids, which can become considerably
stretched in the azimuthal direction as a result of the accretion process
in the disc. Furthermore, the plasmoids produced in the simulations
attain after a few crossing times $t_g:=r_g/c:=GM/c^3$ -- where $G$ is
the gravitational constant, $M$ the mass of the black hole, $c$ the speed
of light and $r_g$ the gravitational radius -- a high azimuthal velocity
which can be well super-Keplerian and have an elongated structure
extending over $\pi/10$ radian. Interestingly, \cite{Matsumoto2020} have
found that a better fit for the observed flare from the galactic source
Sgr~A* is obtained when modelling the circular orbit via a super-Keplerian
motion.

The structure of the paper is the following: In Section \ref{sec:main} we
present the details of our simulations, with subsection \ref{sec:setup}
presenting the numerical setup is discussed and subsection \ref{sec:acc}
reporting the details of the accretion process. The results of the
simulations are instead discussed in Section \ref{sec:results}, with
subsection \ref{sec:Bevo} detailing the evolution of the disc and the
magnetic field, while subsections \ref{sec:plasm} and \ref{sec:plasmoids}
report on the analysis of the plasmoid-formation processes and on their
dynamics, respectively. Finally, we conclude and provide a brief summary
discussion of our results in Section \ref{sec:con}.

\section{Magnetic reconnection in 3D accretion flows: setup} 
\label{sec:main}
%

%
\subsection{Numerical setup} 
\label{sec:setup}

The numerical setup consists of a Kerr black-hole spacetime and of an
initially perturbed torus seeded with a poloidal magnetic field. All
simulations presented here are performed in three spatial dimensions
employing the general-relativistic MHD code \bhac \citep{Porth2017},
which uses second-order shock-capturing finite-volume methods and has
been employed in several investigations \citep{Nathanail2018c,
  Mizuno2018, Nathanail2020b}. The code uses the constrained-transport
method \citep{DelZanna2007}, in order to preserve a divergence-free
magnetic field \citep{Olivares2019}, and has been thoroughly tested and
compared with GRMHD codes of similar capabilities \citep{Porth2019}.

Our initial data consist of an equilibrium torus with constant specific
angular momentum $\ell=4.28$ ~\citep{Fishbone76}, orbiting around a Kerr
black hole with three dimensionless spins, \ie $a = 0,\, 0.5$ and
$0.95$. The inner and outer radii of the torus are at $r_{\rm in}=6\,
r_g$ and $r_{\rm out}=12\,r_g$ respectively, whereas the outer boundary
of the computational domain lies at $500\,r_g$.

As mentioned above, initially the torus is seeded with magnetic field
which is purely poloidal, and consists of a series of nested loops with
varying polarity (\ie the magnetic field in neighbouring loops has a
clockwise direction, which is then followed by a counter-clockwise one
when seen in a vertical cross-sectional plane).
The respective vector potential has the form:
\begin{align}
  &A_{\phi}\propto \max\left(\frac{\rho}{\rho_{\rm max}}-0.2,0\right)\cos((N
  -1)\theta) \sin\left(\frac{2\pi \, (r-r_{\rm in} )}{\lambda_r}\right) 
  \,,
\end{align}
where the maximum rest-mass density in the torus is denoted with
$\rho_{\rm max}$ and the additional parameters ($N\geq 1$ and
$\lambda_r$) set the number and the characteristic length-scale of the
initial loops in the torus. For the results presented here, we consider
only models with $N=3$ and $\lambda_r=2$. Similar magnetic-field
configurations have been thoroughly presented and analysed in the
force-free limit in 2D and 3D \citep{Parfrey2015, Yuan2019, Yuan2019b,
  Mahlmann2020}.

The computational domain uses spherical logarithmic Kerr-Schild
coordinates covering $r\in [0.88 r_g, 500r_g]$, $\theta\in[0,\pi]$ and
$\phi\in[0,2\pi]$ with a base resolution of $N_r \times N_{\theta} \times
N_{\phi}= 192 \times 192 \times 96$ cells. By exploiting \texttt{BHAC}'s
adaptive mesh-refined capabilities, two additional sub-grids are
introduced in those regions where magnetic reconnection is expected to be
efficient, and plasmoid production occurs. As a result, a second level is
placed within a radius of $3 < r/r_g <30$ and a third level is placed at
$5 < r/r_g <20$. Compared to the reference case from \cite{Porth2019},
which simulates a larger radial extent of $2500\,r_g$, the resulting
effective resolution in the regions of prime interest is hence $N_r
\times N_{\theta} \times N_{\phi}= 1050 \times 768 \times 384$.

\begin{table}
  \centering
   \caption{Initial parameters for the models considered. The second
     column refers to the dimensionless spin of the black hole, whereas
     the third to the initial average magnetisation in the torus. The
     last column corresponds to the module used in \bhac Ideal or
     Resistive. Each model name (in the first column) encodes information
     of the model parameters.}
	\begin{tabular}{l|c|c|l|} 
	\hline \hline 
	model  & $a$& $\langle\sigma_{\rm init}\rangle$ & module \\
	\hline
	\texttt{HB.id.0.0}  & $0.0$ & $10^{-5}$ & Ideal\\ 
	\texttt{HB.id.0.5}  & $0.5$ & $10^{-5}$ & Ideal\\ 
	\texttt{LB.id.0.0}  & $0.0$ & $10^{-4}$ & Ideal\\ 
	\texttt{HB.res.0.9} & $0.9$ & $10^{-5}$ & Resistive \\ 
        \texttt{HB.id.0.9}  & $0.9$ & $10^{-5}$ & Ideal\\
	\hline \hline 
	\end{tabular}
\label{table:models} 
\end{table}

For the three ideal-GRMHD models (see Table \ref{table:models}),
dissipation of the magnetic energy is entirely numerical and, as was
shown in 2D simulations, physically meaningful results can nevertheless
be obtained when exploring the behaviour of the solution with increasing
resolution \citep{Obergaulinger:2009, Rembiasz2017, Nathanail2020,
Obergaulinger2020}. For the simulation run that has physical resistivity,
\ie for model \texttt{HB.res.0.9}, we solve the resistive GRMHD equations
in the version implemented in \texttt{BHAC} \citep{Ripperda2019}, and we
assume a uniform and constant resistivity with a rather small value of
$\eta =5\times 10^{-5}$. The small resistivity is chosen in order to
reproduce a near-ideal limit for the accretion flow dynamics, but also to
allow for a physical reconnection that will result in plasmoid formation
\citep{Ripperda2019b, Ripperda2020}. When assuming the gravitational
radius, $r_g$, as the typical lengthscale, the chosen resistivity results
in a large lundquist number of $S := r_g c/\eta = 10^4$, that is, the
minimum lundquist number needed to produce plasmoids under the physical
conditions considered here \citep{Bhattacharjee2009,
Uzdensky2010,Ripperda2019b,Ripperda2020}.

\subsection{Properties of the accretion flow}
\label{sec:acc}

\begin{figure}
  \begin{center}
    \includegraphics[width=0.95\columnwidth]{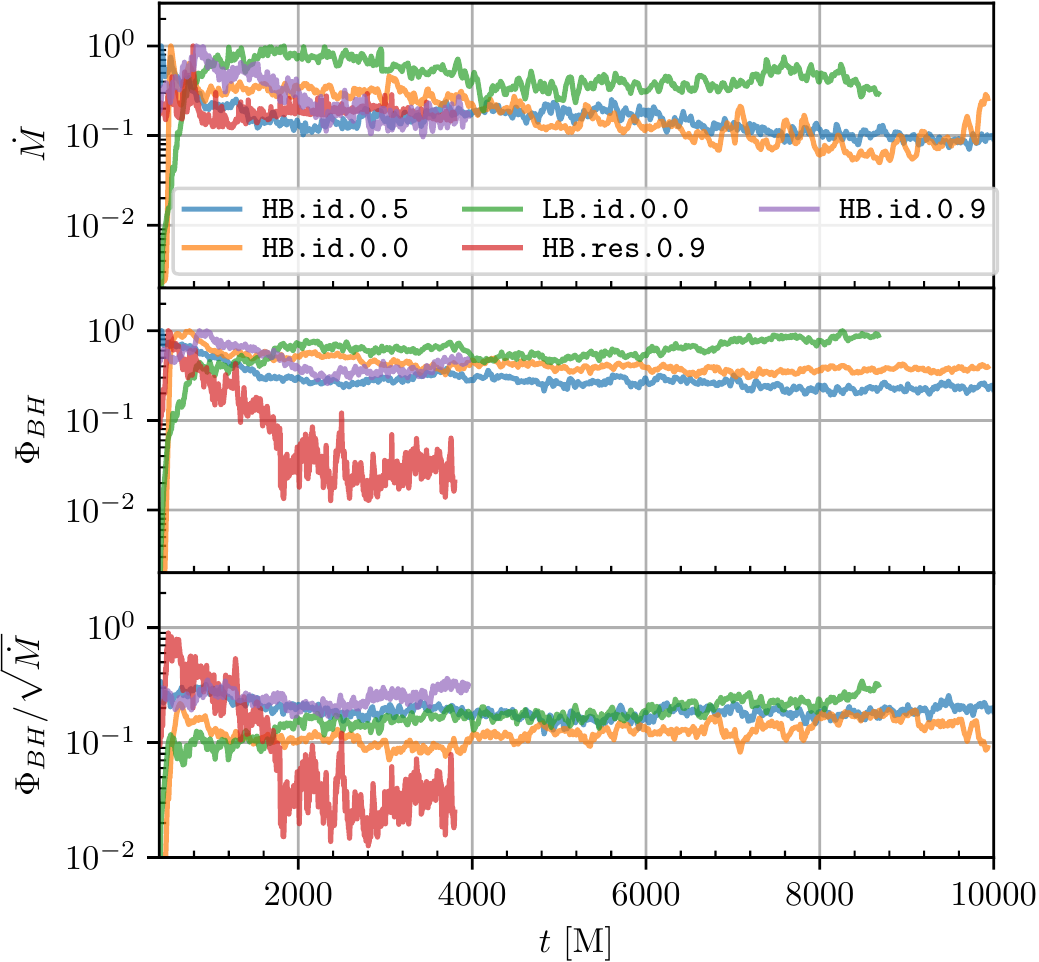}
  \end{center}
  \caption{Upper panel: evolution of the mass-accretion measured across
    the black hole event horizon. Middle panel: evolution of the magnetic
    flux accreted onto the black hole. Lower panel: evolution of the
    normalized magnetic flux accumulated on the black-hole
    horizon. Different lines refer to the different models considered and
    summarised in Table \ref{table:models}.}
    \label{fig:mdot}
\end{figure}

The main focus of our work is the magnetic reconnection occurring close
to event horizon and the production of plasmoids through this process.
However, we need to analyze the main properties of the accretion dynamics
through the mass-accretion rate and the magnetic flux across the
horizon. We measure the former as
\begin{align}
 \dot{M}:=\int_0^{2\pi}\int^\pi_0\rho u^r\sqrt{-g}\,d\theta d\phi\,,
\label{mdot}
\end{align}
and we report its behaviour as function of time in the upper panel of
Fig. \ref{fig:mdot} for all of the tori considered. Note that, after a
time of $\sim 3000\,M$, the accretion process reaches a quasi-steady
state till the end of the simulation for all models, thus highlighting
that a stationary turbulent state -- produced by the development and
saturation of the magnetorotational instability (MRI) -- is sustained in
the torus.

\begin{figure}
  \begin{center}
    \includegraphics[width=0.95\columnwidth]{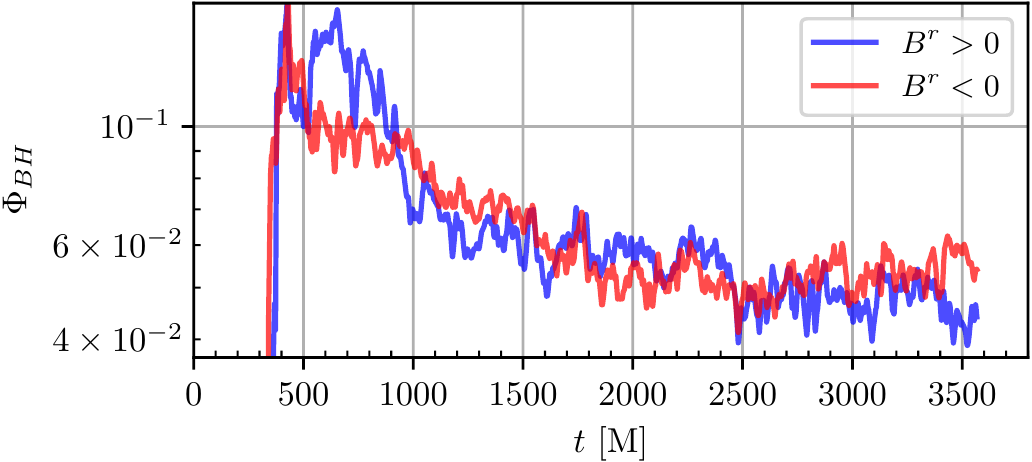}
  \end{center}
  \caption{Evolution of the magnetic flux accreted onto the black hole
    over one hemisphere for model \texttt{HB.id.0.5}. The blue shaded
    line corresponds to a magnetic flux of positive polarity (\ie the
    integrated flux with $B^r>0$) and red dashed line to the other
    polarity (\ie integrated flux with $B^r<0$).}
    \label{fig:phi+-}
\end{figure}

\begin{figure*}
  \begin{center}
    \includegraphics[width=0.95\textwidth]{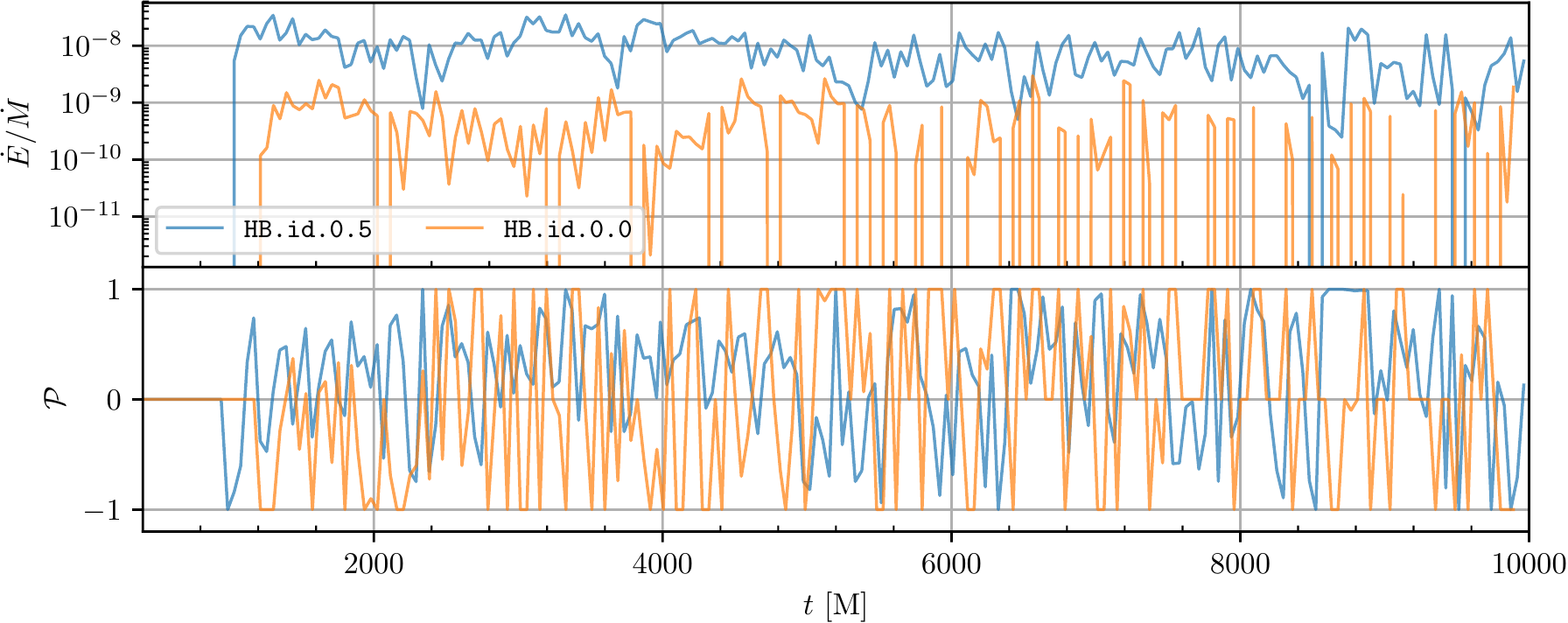}
  \end{center}
  \caption{Upper panel: evolution of the power of the outflow when
    normalized by the mass-accretion rate; note that the energy of the
    outflow shows large fluctuations and for nonrotating black holes
    (model \texttt{HB.id.0.0}) these are of more than two orders of
    magnitude larger than for rotating ones. Lower panel: evolution of
    the north/south asymmetry in the jet as defined in
    Eq. \eqref{pjetns}; note that for all models there is no steady
    emission from the upper or the lower part of the outflow. }
    \label{fig:eta}
\end{figure*}

\begin{figure}
  \begin{center}
    \includegraphics[width=0.49\textwidth]{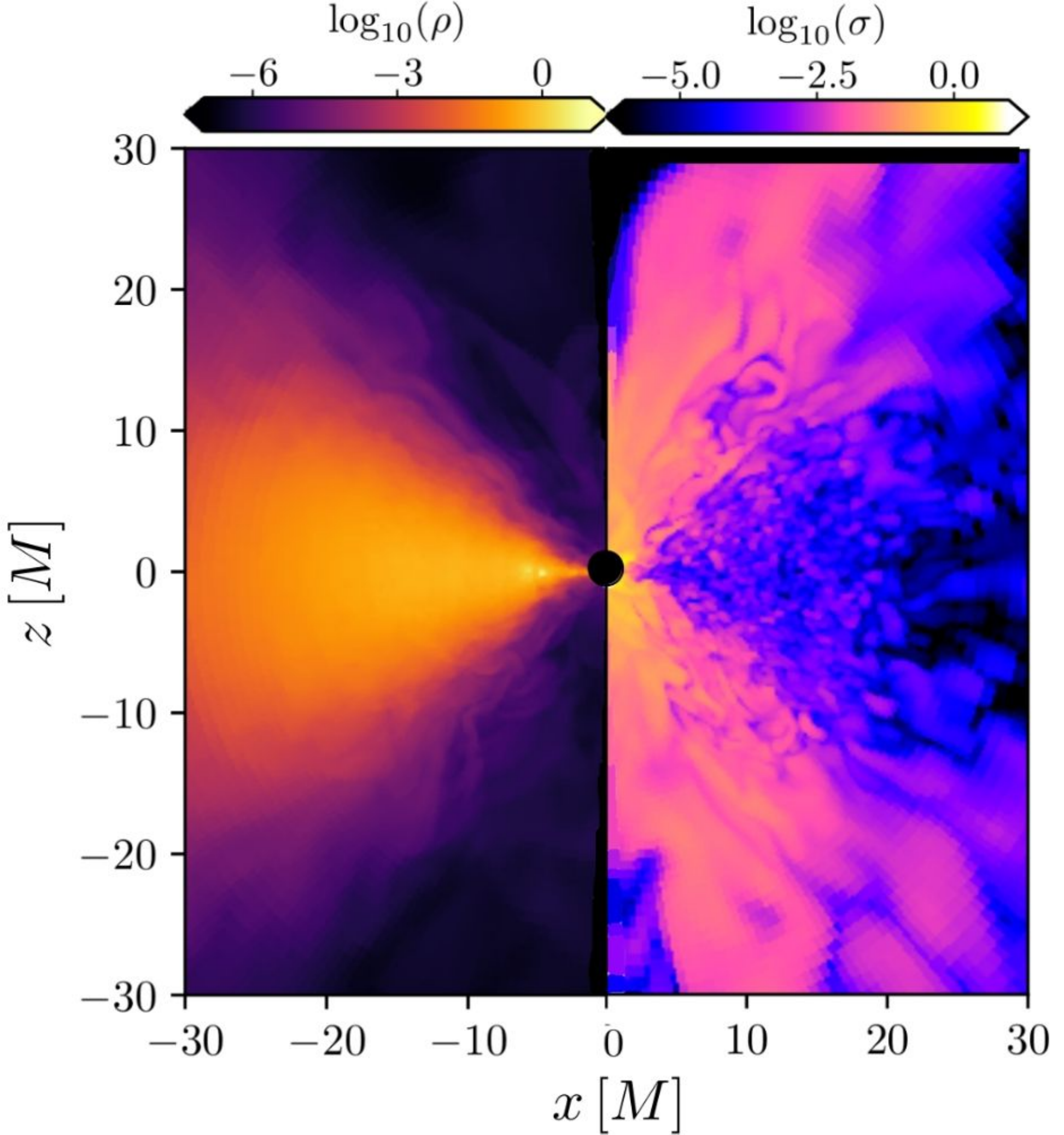}
  \end{center}
\caption{Distribution of the rest-mass density (left part) and of the
  magnetisation (right part) for model \texttt{HB.id.0.5}. Note the
  typical presence of a low-density funnel where the magnetisation can
  acquire higher values.}
    \label{fig:Blines_a}
\end{figure}

\begin{figure*}
  \begin{center}
    \includegraphics[width=0.70\textwidth]{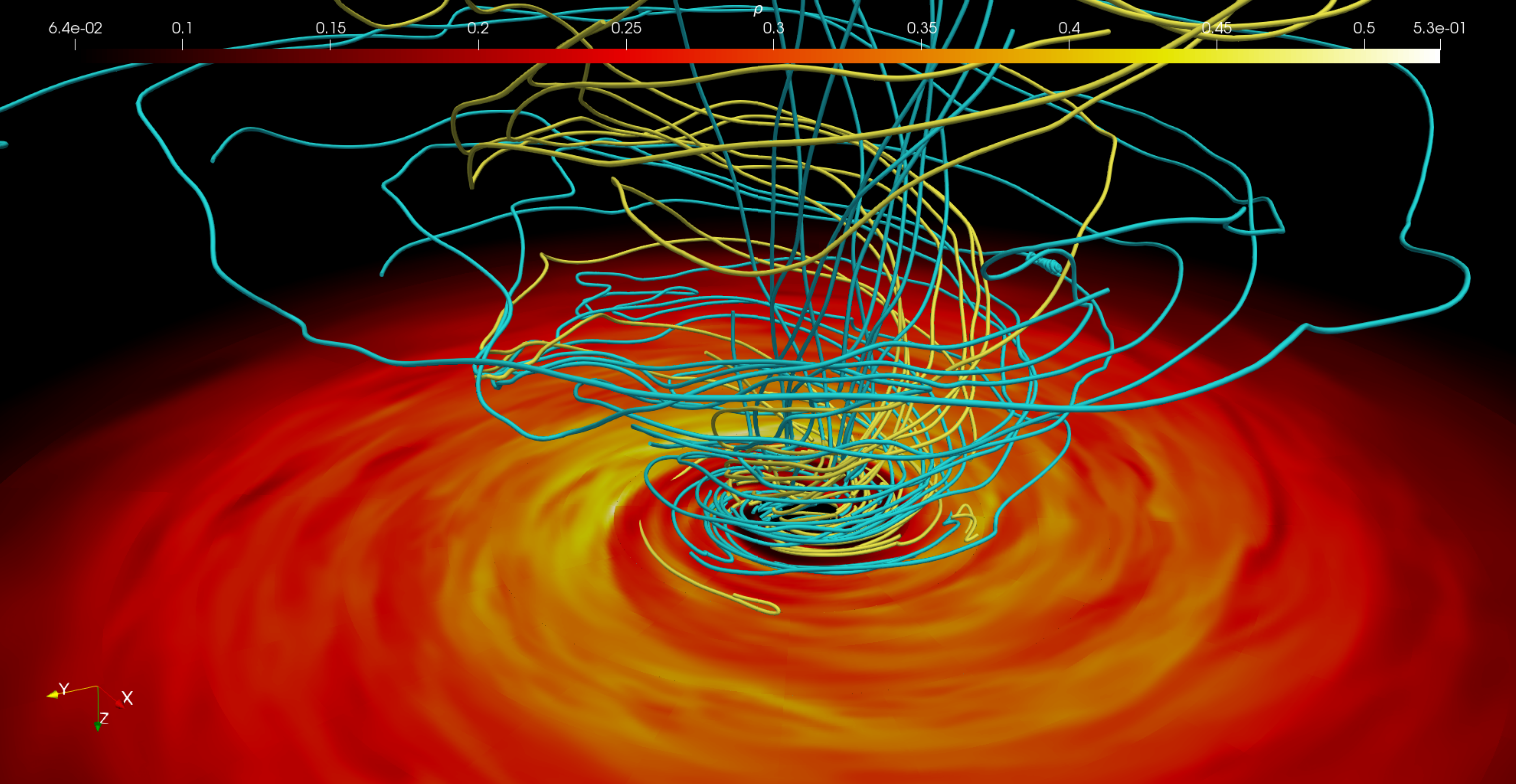}
  \end{center}
\caption{Three-dimensional representation of the accretion flow and of
  the magnetic-field topology. The rest-mass density distribution is
  shown only in the equatorial plane with a yellow-red colorcode. Shown
  instead with cyan and yellow colours are the magnetic-field lines of
  different polarity that will lead to reconnection and plasmoid
  formation in the funnel. The data refers to the rotating black-hole
  model \texttt{HB.id.0.5}. }
    \label{fig:Blines_b}
\end{figure*}


The magnetic flux accreted across the event horizon is instead defined as
\begin{align}
  \Phi_{\rm BH}:=\frac{1}{2}\int_0^{2\pi}\int^\pi_0 |B^r|\sqrt{-g}d\theta d\phi\,,
\label{phiBH}
\end{align}
while the ``normalized'' magnetic flux $\phi_{\rm BH}:=\Phi_{\rm
  BH}/\sqrt{\dot{M}}$ is shown in the lower panel of Fig. \ref{fig:mdot}
for all of the models considered. To better distinguish the evolution of
the magnetic flux of different polarity, we perform the integrals by
considering separately the positive and the negative $B^r$ component in
one hemishpere. This is shown in Fig. \ref{fig:phi+-}, which reports how
there is no magnetic flux threading the horizon initially. However, as
accretion is triggered, both polarities are dragged in and the flux of
positive $B^r$ (blue dashed line) becomes slightly larger than the
corresponding negative one. However, as the evolution procees, also the
flux of negative $B^r$ (red dashed line) increases, reaching levels that
are comparable to the positive one. Overall, once stationarity is
reached, the two fluxes are fluctuating but comparable, as one would
expect in a situation where large amounts of reconnection takes place.

We recall that the most extensively investigated modes of disc accretion
onto a black hole are either of SANE (Standard And Normal Evolution)
\citep{Narayan2012}, or MAD (Magnetically Arrested Disc)
\citep{Igumenshchev2003} type. Among the various differences in their
properties \citep[see, \eg][]{EHT_M87_PaperV}, possibly the most salient
one is given by the typical value of the normalised flux, which is around
$\phi_{\rm BH}=\phi_{\rm max}\approx 15$ (within the units adopted here
which matches the limiting flux quoted by \citet{Tchekhovskoy2011}) for a
MAD accretion. In particular, a MAD state can be obtained when the
accretion of magnetic flux onto the event horizon is sufficiently high to
reach equipartition with the disc ram pressure and possibly even counter
accretion \citep{Igumenshchev2003}. In this respect, the models
considered here do not acquire a significant magnetic flux during their
evolution and hence cannot be considered as MAD accretion discs. On the
other hand, they cannot be considered SANE accretion discs either, since
these are produced with tori having nested loops of one polarity and are
therefore characterised by magnetized funnels with a much lower
normalized magnetic flux and a much lower jet power
\citep{EHT_M87_PaperV, Wong2021}. The initial magnetic-field
configuration and the size of the torus are the most important parameters
for the resulting evolution. A large-scale toroidal magnetic field, with
a rather high initial magnetization, has been shown to produce powerful
jets \citep{Christie2019b}. During such evolutions, however, the global
large-scale magnetic polarity can change sign and so does the magnetic
flux across the event horizon. This behaviour is different from what
reported here, where the magnetic flux across the event horizon can
change sign (even on small timescales; see Fig.~\ref{fig:phi+-}), but
where the global large-scale magnetic field always admits both
polarities, which are therefore susceptible to intense reconnection. A
change of sign of the radial magnetic field component in the evolution of
the accretion flow was observed in \citet{Barkov2011}, where the initial
magnetic-field configuration had two large magnetic loops of different
polarity. From the analysis reported by \citet{Barkov2011}, it seems that
first a jet of one polarity was established and then, as magnetic flux of
different polarity was accreted, the jet was destroyed and
re-established.

We believe the reason why our accretion discs do not fall in either of
the two classes is because or initial magnetic-field topology results in
a fluctuating accretion of magnetic flux. More specifically, magnetic
flux of one polarity is brought to the event horizon of the black hole,
and as the magnetic flux of the opposite polarity reaches the vicinity of
the horizon, they annihilate reducing the overall flux across the event
horizon. In this way, no stable jet can be produced and indeed this what
the simulations reveal. In addition, another distinctive feature of our
accretion flows is that there is no need to reset to atmosphere values in
the regions within the funnel. This is because the funnel is neither
depleted of plasma, nor reaches high values of magnetisation, as instead
happens in the case of MAD ans SANE accretion discs.

We measure the energy released from the outflow through the energy flux
that passes through a 2-sphere placed at $r=50 \,r_g$ and compute it as
\begin{align}
\dot{E}:=\int_0^{2\pi}\int^\pi_0 (-T^r_t -\rho u^r)\sqrt{-g}d\theta
  d\phi\,,
\label{pjet}
\end{align}
the integrand in Eq. \eqref{pjet} is set to zero when surface $\sigma
\leq 1$ \citep{Porth2019}. The energy of the outflow in code units and
normalized with the mass-accretion rate is shown in the upper panel
Fig. \ref{fig:eta}. Note that the evolution of the energy of the outflow
shows large fluctuations and for some models (\eg \texttt{HB.id.0.0})
these fluctuations are of more than two orders of magnitude larger. To
better understand which part of the outflow is giving most of the energy,
we explore the jet and counter jet (upper/lower) asymmetry measuring the
asymmetry of the outflow and the contribution from the upper and lower
(above and below the black hole) via the parameter $\mathcal{P}$ defined
as follows \citep{Nathanail2020} 
\begin{align}
	\mathcal{P}:= \frac{\dot{E}_{\rm u}- \dot{E}_{\rm
	l}}{\dot{E}_{\rm u} + \dot{E}_{\rm l}}\,,
\label{pjetns}
\end{align}
where $\dot{E}_{\rm u}$ and $\dot{E}_{\rm l}$ are the power of the upper
and lower part of the outflow, respectively. Clearly, when the upper jet
dominates $\mathcal{P} \simeq 1$, and when the lower jet dominates
$\mathcal{P} \simeq -1$, while when $\mathcal{P} \simeq 0$ the jets are
symmetric [we set $\mathcal{P}=0$ whenever $\sigma < 1$ in expression
  \eqref{pjet} 

The inspection of the lower panel of Fig. \ref{fig:eta} reveals that for
all models there is no steady emission from the upper or the lower part
of the outflow. More specifically, the zero spin model \texttt{HB.id.0.0}
switches its emission in the upper and lower vertical directions with
higher periodicity than the spinning $a=0.5$ model
\texttt{HB.id.0.5}. Furthermore, for the $a=0$ model \texttt{HB.id.0.0},
a low jet power is measured and indeed expected on the basis of the very
small values of the magnetic flux accreting onto the black hole in this
case. Note that for clarity Fig. \ref{fig:eta} reports the behaviour of
only two representative models but the behaviour illustrated is shared
across all of the models considered.

\begin{figure*}
  \begin{center}
    \includegraphics[width=0.95\textwidth]{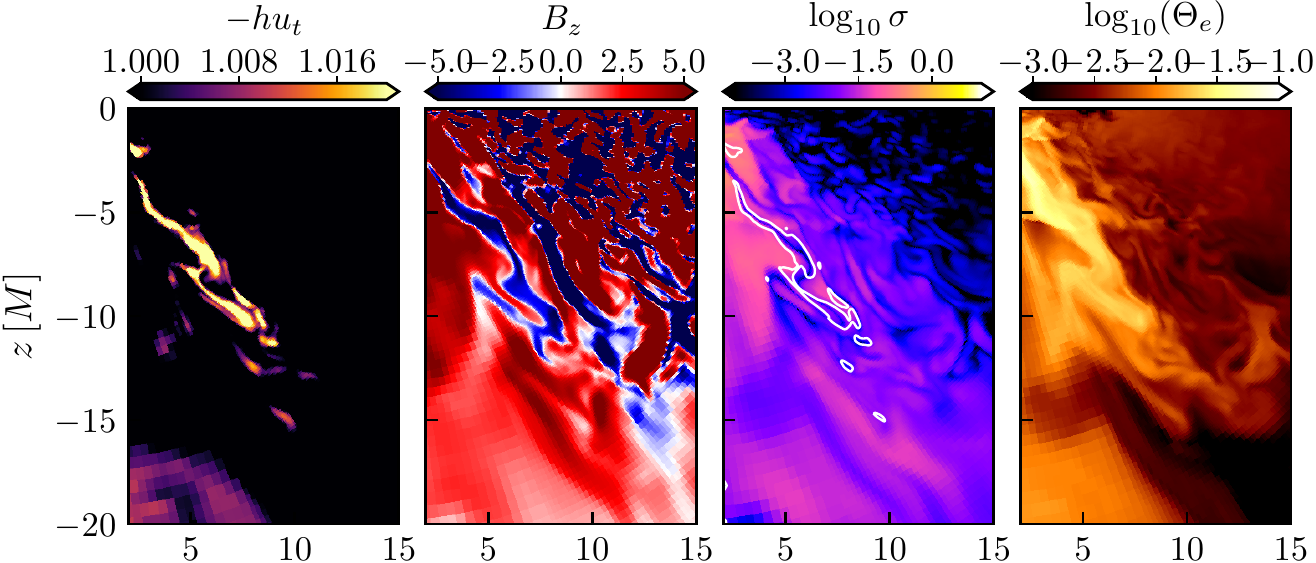}
    \includegraphics[width=0.95\textwidth]{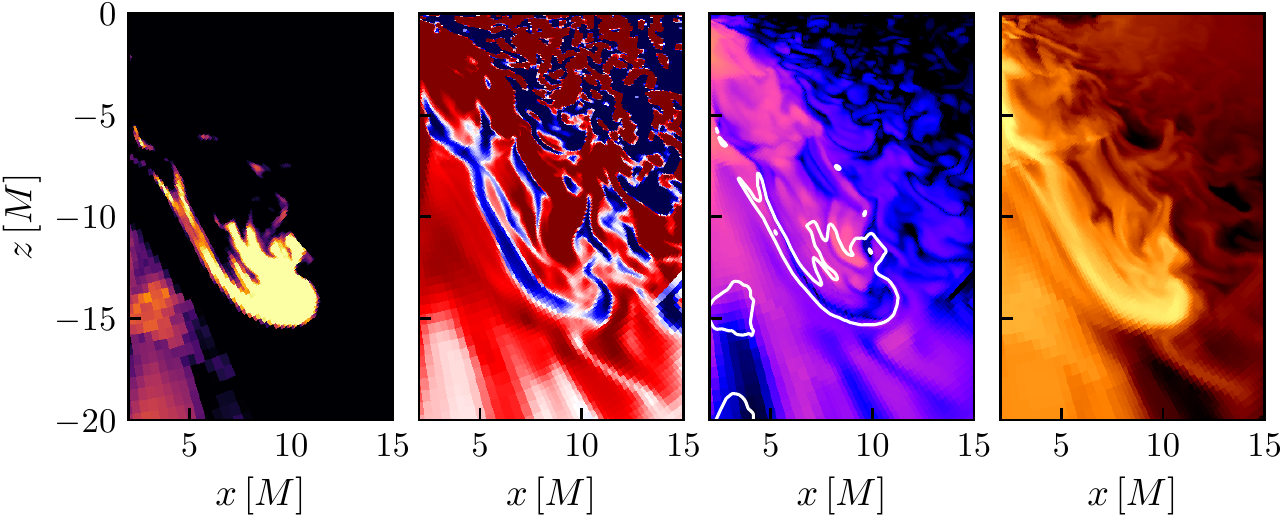}
  \end{center}
\caption{Two-dimensional section at $\phi=0^{\circ}$ showing the production
  of a plasmoid. From left to right, the four panels report: the Bernoulli
  parameter $-hu_t$, the vertical magnetic-field component $B^z$, the
  magnetisation scalar $\sigma$, and the dimensionless temperature
  $\Theta_e$. The data refers to the ideal-MHD model \texttt{LB.id.0.0}
  at time $t=2860 \, M$ in the upper panels and at $t=2950 \, M$ in the
  lower panels.}
    \label{fig:sigma}
\end{figure*}
%

\section{Magnetic reconnection in 3D accretion flows: results} 
\label{sec:results}

%
\subsection{Disc and Magnetic field evolution}
\label{sec:Bevo}

VLBI (Very Large Baseline Interferometry) imaging of supermassive black
holes on scales comparable with that of the event horizon requires a
detailed knowledge of the physical environment near the black hole and is
essential in distinguishing between theoretical models of the plasma
condition and different types of compact objects \citep{Mizuno2018,
  EHT_M87_PaperV, EHT_M87_PaperVI, Olivares2020}. Similarly, polarization
measurements can give more insight in the magnetic field structure and
its activity, together with microphysics of emission, such as the
production of an electron population with a non-thermal energy
distribution \citep{EHT_M87_PaperVII, EHT_M87_PaperVIII, Goddi2021,
  Mizuno2021}. Finally, probing the topology of the magnetic field can be
also essential for understanding flaring activity \citep{Abuter2018b}.

Our simulations reveal that, as the accretion onto the black hole tends
to a quasi-stationary state, the magnetisation tends to rise
significantly in the low-density funnel region and provides conditions
that are favourable for an efficient plasmoid production. This can be
seen in Fig. \ref{fig:Blines_a}, where we report the rest-mass density
(left part) of the matter both in the torus (where is very high) and in
the funnel region, where it is considerably reduced. Also shown with a
different colorcode is the magnetisation (left part), which is instead
very large in the funnel.

The central black hole absorbs the plunging matter that has arrives on
the event horizon via the development of the MRI and in doing so has
advected large amounts of magnetic field. Inside the torus, the magnetic
field is mostly chaotic as a result of the highly turbulent motion there;
however, as it reaches the funnel region, it becomes more and more
twisted due to the disc rotation and, in the case of rotating black
holes, also the spacetime rotation induced by the dragging of inertial
frames. However, even for black holes with a zero spin, a similar
magnetic field topology is observed, since the field lines are twisted
mostly because of the motion of the fluid.

Thanks to our initial magnetic-field with poloidal loops of alternating
polarity, a very specific magnetic-field configuration is produced in the
funnel region, as shown in Fig. \ref{fig:Blines_b}, where magnetic-field
lines -- mostly of poloidal nature -- and of different polarity are
depicted with cyan and orange colour, respectively. These field lines are
rooted near the event horizon and it is clear that the three-dimensional
interfaces where these lines intersect in the high-magnetisation funnel
are potential reconnection sites. Configurations of this type could be
relevant for production of striped jets -- that is, a jet with
alternating toroidal magnetic-field polarity along the propagation axis
-- especially if the initial magnetic field is strong and thus leading to
more powerful outflows \citep{Giannios2019}. More importantly, the
reconnection sites produced by this magnetic-field configuration with
poloidal magnetic fields having alternating polarities are quite
different from the reconnection sites found at the funnel wall, which
have lower magnetisation -- both for SANE and MAD accretion modes -- or
from the reconnection taking place on the equatorial current sheet.
Indeed, in typical single-loop initial magnetic fields, plasmoid
production occurs almost exclusively at the edge of the jet funnel, that
is, the jet sheath and not inside the funnel, which is instead filled
with highly magnetized material and poloidal magnetic-field lines of the
same polarity that produce a powerful outflow.

\begin{figure*}
  \begin{center}
    \includegraphics[width=0.95\textwidth]{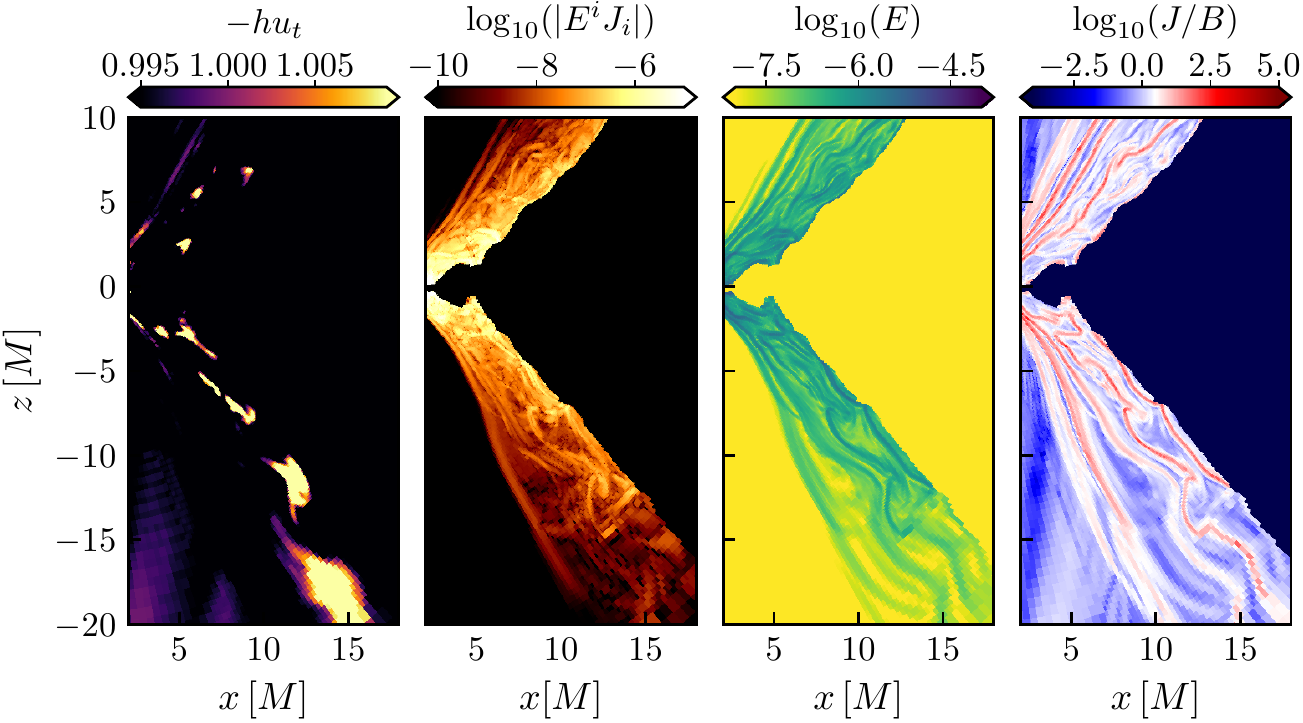}
  \end{center}
\caption{Two-dimensional section at $\phi=0^{\circ}$ showing the
  production of a plasmoid. From left to right, the four panels report:
  the Bernoulli parameter $-hu_t$, the magnitude Ohmic heating
  $\log_{10}(|E^iJ_i|)$, the electric-field strength $\log_{10}(E)$, and
  the the ratio between the electric current density and the
  magnetic-field strength $\log_{10}(J/B)$. The data refers to the
  resistive-MHD model \texttt{HB.res.0.9} at time $t=1130 \, M$; note
  that the high-density region in the torus is masked.}
    \label{fig:res}
\end{figure*}

In Fig. \ref{fig:Blines_b} we have masked the torus above and below the
equatorial plane so as to better visualise the magnetic-field topology,
leaving the rest-mass density visible only near the equatorial plane in
order to illustrate the motion of the fluid. This representation clearly
highlights that determining the launching sites of magnetic reconnection
have an intrinsically 3D nature and that any 2D representation would
inevitably provide only a limited view, albeit still correct
\citep{Nathanail2020, Ripperda2020}. This limitation of 2D simulations
becomes particularly severe in those conditions -- such those encountered
in our simulations -- where the accretion flow is particularly variable
and chaotic.

\subsection{Plasmoids: physical conditions at generation}
\label{sec:plasm}

Current sheets are formed everywhere inside the torus due to the
turbulent nature of the magnetic field there. However, because of the
large rest-mass density and the small magnetisation in the torus, the
efficiency of magnetic reconnection in such regions is rather low and
hence reconnection does not heat the local plasma significantly. On the
other hand, the efficiency of magnetic reconnection is expected to be
rather high at the funnel sheath, where the rest-mass density decreases
and there is a significant gradient in the magnetisation, which increases
moving towards the polar axis.

Inside the funnel, the magnetic-field lines are sheared and twisted (as
seen in Fig. \ref{fig:Blines_b}) due to differential rotation of the
accreting fluid. Reconnection of these flux tubes releases magnetic
energy that heats up the plasma close to relativistic temperatures, with
values of the dimensionless temperature that can be as large as
$\Theta_e:=p/\rho\approx 0.1$ in those regions where plasmoids are
generated. Furthermore, a portion of the plasmoids produced in this way
can reach energies sufficiently large to become gravitationally unbound,
although this does not necessarily imply that the plasmoids will actually
reach spatial infinity as some them are actually moving inwards and are
therefore accreted by the black hole.

In Fig. \ref{fig:sigma} we present some representative snapshots
portraying the production of a large-scale plasmoid from model
\texttt{LB.id.0.0} when represented in a 2D slice at $\phi=0$. The upper
and lower rows refer respectively to times $t =2860\,M$ and $t= 2950\, M$
and have been selected to illustrate how a time interval of less than
$100\,M$ is sufficient to build a large plasmoid with a radius of
$\approx 3-5\,r_g$. Note that the reconnecting current sheets produce
filaments of small plasmoids -- or plasmoid chains -- which, over time,
merge leading to a single large plasmoid that spirals and moves outwards
in this case.

\begin{figure*}
  \begin{center}
	  \includegraphics[width=0.70\textwidth]{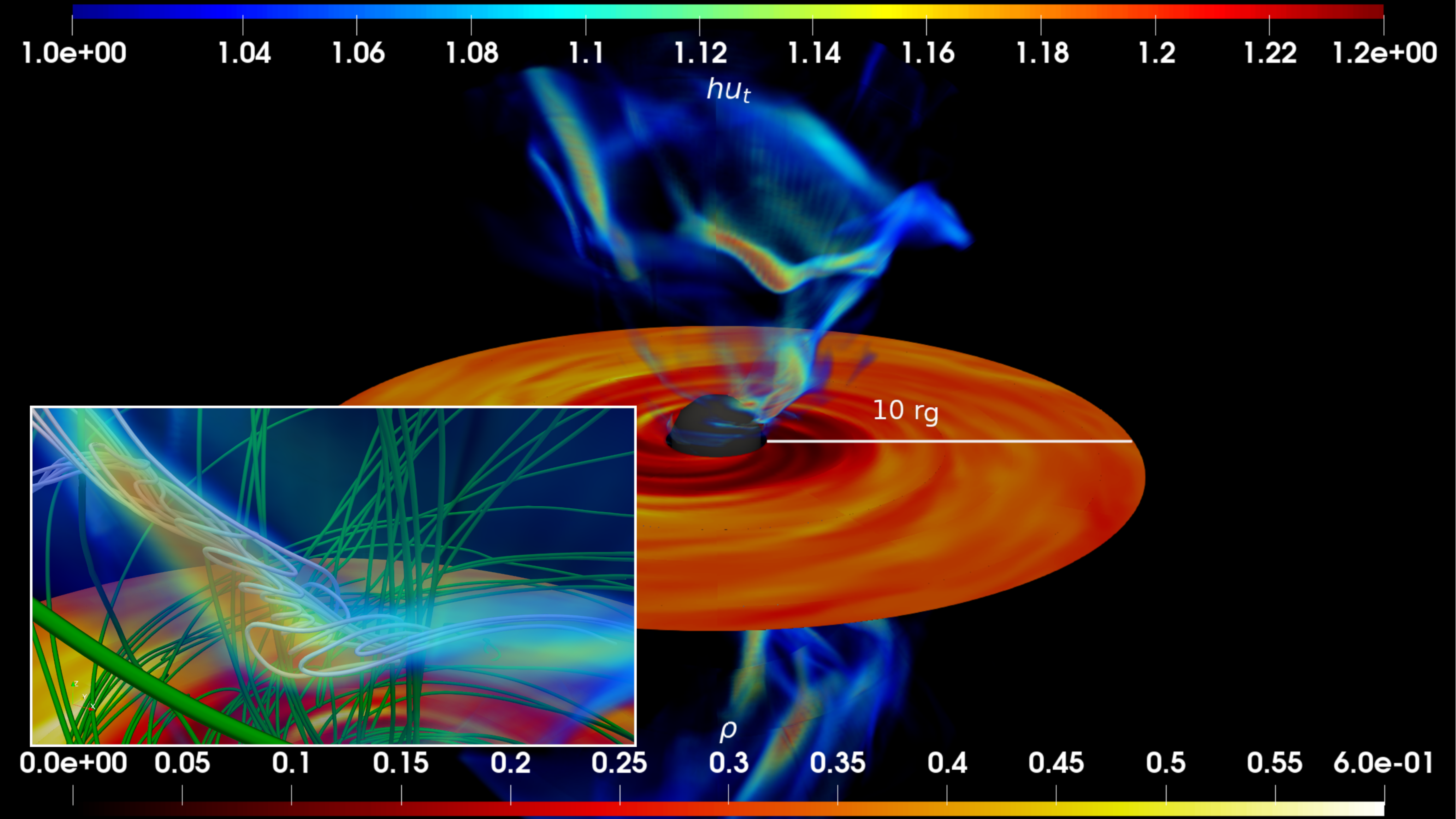}  
  \end{center}
  \caption{Three-dimensional view of the magnetisation for model
    \texttt{LB.id.0.0} and at time $4200\, M$. Note that the rest-mass
    density is shown only on the equatorial plane so as to have a sense
    of the rotation in the torus, The colourcode for the magnetisation is
    chosen to range between 1.00 and 1.20, so that those regions with
    $\sigma \gtrsim 1.15$ effectively represent a plasmoid. The small box
    offers a zoomed view of the largest plasmoid of the snapshot;
    magnetic field outside the plasmoid are shown in green, while those
    inside the plasmoid as shown in cyan.}
    \label{fig:3Dplasm}
\end{figure*}

More specifically, the first column of panels in Fig. \ref{fig:sigma}
show the Bernoulli parameter $hu_t$ at different times -- where $h$ is
the specific enthalpy and $u_t$ the time covariant component of the fluid
four-velocity \citep{Rezzolla_book:2013}. When $hu_t < -1$, this quantity
can be used distinguish fluid elements that are gravitationally unbound
as a result of the energization through reconnection, and which can in
principle leave the black hole. The second column of panel shows instead
the vertical component of the magnetic field, $B^z$, using a colourcode
that emphasises the different polarities. In this way, it is possible to
highlight the reconnecting current sheet on the left part of the plot,
where the layer of blue regions (negative polarity) lies inside the red
area (positive polarity). In the third column we report the magnetisation
parameter $\sigma$, which shows that exactly at the location of the
current layer, the magnetisation has decreased due to occurrence of
reconnection. Finally, the last column presents the dimensionless
temperature $\Theta_e$, highlighting how this increases as a result of
reconnection. We note that reconnection with mild magnetisation, $\sigma
\approx 0.1-1$ results in a significant heating of the plasmoid, with
relativistic temperatures $\Theta_e\approx 0.1$. Such plasmoids can power
episodic outflows as the accretion takes place.

A rather important aspect of the plasmoid-formation process that we have
not yet discussed about is the presence of a considerable azimuthal
motion in the plasmoids formed in our simulations. Hence, in addition to
the intrinsic and background azimuthal motion of the accreting plasma
\citep{Shende2019}, plasmoids produced in 3D simulations can acquire an
additional and considerable acceleration in the azimuthal direction when
produced via reconnection. It is important to note here that plasmoids
are significantly affected by the local guide field \citep{Edmondson2017}
and, indeed, episodic magnetic reconnection, such as break-out
reconnection, has been observed in the solar corona
\citep{Kumar2019plasm}. A similar process can be thought to take place
also in this accretion scenario, as the plasma enters the funnel region
and reconnecting current sheets can efficiently accelerate plasmoids and
generate particles with energy distributions that are non-thermal and
which we do not model here.

We have already introduced that reconnection layers with high lundquist
numbers are unstable to the plasmoid instability and result in fast
magnetic reconnection with a high reconnection rate, which becomes
essentially independent of the value of the plasma resistivity
\citep{Bhattacharjee2009, Uzdensky2010, Stanier2019, Ripperda2020}. This
process -- and in particular the weak dependence of the reconnection rate
on the resistivity once a sufficiently high lundquist number has been
achieved -- has been shown to be present also in the case of turbulent
reconnection \citep{Lazarian2020}. In our resistive model
\texttt{HB.res.0.9}, we find that the lundquist number is high enough,
namely $S\gtrsim 10^4$, that the plasmoid instability is triggered. In
this model, for which we solve the equations of GRRMHD, magnetic
reconnection and energy dissipation is a result of physical
resistivity. In this way, we can we analyse in a consistent (although not
necessarily realistic) manner the Ohmic heating taking place during the
reconnection events and also monitor the properties and evolution of the
electric field. Moreover, through this model, we can perform a close
comparison between the ideal-MHD and the resistive-MHD simulations we
have performed.

Figure \ref{fig:res} reports a plasmoid chain produced at $t= 1130\,M$ in
the resistive model \texttt{HB.res.0.9}. In the first panel we show again
the Bernoulli parameter $hu_t$ and in the in subsequent panels we mask
the high-density region of the torus in order to highlight the important
regions outside the torus where magnetic reconnection is most efficient.
In particular, in the second panel we show in a logarithmic scale the
magnitude Ohmic heating $E^iJ_i$, where $E^i$ and $J^i$ are the
components of the electric field and current, respectively. Note that the
Ohmic heating is significant at the layer entering the funnel and peaks
close to the black hole event horizon. Furthermore, this strong parallel
electric field can efficiently accelerate particles close to the event
horizon, but also along the sheath of the funnel region which has a width
of $\approx 5\,r_g$. As a result of these accelerations, the energy
distribution of the electrons involved can deviate significantly from a
thermal distribution \cite{Mizuno2021}.

The third panel, on the other hand, shows the magnitude of the electric
field, which clearly peaks there where the current-sheet layers
develop. Finally, the fourth panel reports the ratio between the electric
current density $J:= \sqrt{J^iJ_i}$ and the magnetic-field strength
$J/B$, that again traces the reconnecting current layers at the sheath of
the funnel region. Note that at the current sheets the electric current
$J$ can be between two and three orders of magnitude larger than the
magnetic field $B$.

Also in the case of the resistive simulation, larger plasmoids with a
width of a few $r_g$ are produced in the over a timescale of about $100
\,M$, as already observed in the ideal-MHD simulations. This allows us to
estimate the reconnection rate to be $\approx 0.01$, which hints to the
occurrence of fast magnetic reconnection consistent with the high
lundquist number associated with our model \citep{Bhattacharjee2009,
  Uzdensky2010, Ripperda2019b}. Finally, the ability to model also a
magnetic field allows us to estimate the inflow velocity of the current
sheets in terms of the radial drift velocity, $(\vec{E}\times
\vec{B})^r/B^2$, for which can can measure in the neighbourhood of the
current sheets a value of $|(\vec{E}\times \vec{B})^r/B^2| \approx
10^{-2}$.

All things considered, the evolution of the resistive model and the
plasmoids formed in the corresponding accretion flow are rather similar
to the ideal runs we have performed. Most probably this is due to the
small resistivity used, which is however essential in order to model the
accretion dynamics accurately and avoid excessive magnetic diffusion in
the torus. As a way to improve our resistive-MHD description, we could
employ a non-uniform resistivity so as to reproduce a physical scenario
consisting of a low-resistivity torus and of a highly resistive funnel
close to the event horizon. An approach following the one suggested by
\citet{Dionysopoulou2015} will be explored in a future work.

\subsection{Plasmoids: launching sites and morphology}
\label{sec:mpla}

We have so far concentrated on the processes giving rise and providing
energy to the plasmoids found in our simulations. Next, we will discuss
in detail where these plasmoids are produced, what are the main features
of their dynamics as they interact with the background shearing flow, and
how they can be related with the flaring activity seen in the
supermassive black hole at the center of our galaxy, Sgr~A*.

Already in Sec. \ref{sec:Bevo} we have illustrated that reconnection is
very active in the low-density funnel as a result of poloidal flux tubes
of alternating polarity that interact and reconnect. Most of the time,
large plasmoids produced in these regions do not fall back to the black
hole, but continue in an outward motion. These plasmoids are different
from those that are produced well inside the torus and whose origin is
linked to the turbulent motion triggered by the development of the
MRI. These other plasmoids, which are created as frequently as they are
destroyed, do not lead to a significant increase of the temperature of
the disc, mostly because of the very low magnetisation in the dense
interior of the torus. A third and final launching site of plasmoids is
represented by the jet sheath, that is, the thin vertical region of rapid
increase of the magnetisation placed between the torus and the jet
interior.

The spatial distribution of the magnetisation can be used to determine
the plasmoid launching sites and deduce their 3D morphology. As already
noted by \citet{Nathanail2020}, the magnetisation is extremely large
within a current sheet (or filament in 2D simulations) although it can
drop to zero at the center of the sheet, where the magnetic field
vanishes and changes sign \citep[see Fig. 8 in][]{Nathanail2020}. In view
of this, we provide in Fig. \ref{fig:3Dplasm}, which provides a 3D view
of the magnetisation for model \texttt{LB.id.0.0} and at time $4200\,
M$. Note that in the figure we have we masked the rest-mass density above
and below the equatorial plane, visualising it only on the equatorial
plane so as to have a sense of the rotation in the torus (the radius of
the disc is $30\,r_g$). The colourcode for the magnetisation is chosen to
range between 1.00 and 1.20, so that those regions with $\sigma \gtrsim
1.15$ can be considered to be representing a plasmoid.

Concentrating on features above the equatorial plane, it is possible to
recognise at least two distinct plasmoids, with an approximate size of
$\sim 10-15\,r_g$ along the filament and a diameter of $\sim
3-5\,r_g$. The first one is near the event horizon and at lower
latitudes, while the second one is further away from the event horizon
and at higher latitudes. Note that their shape is far from spherical and
both plasmoids appear mostly as filaments and with an orientation that is
either mostly poloidal (first plasmoid) or a combination of poloidal and
toroidal (second plasmoid). Hence, the morphology of plasmoids in 3D
simulations is rather different from the one that can be deduced from 2D
simulations, where they appear mostly as spherical. In particular, the 3D
rendering allows one to appreciate that the shearing present in the
background flow inevitably introduces a toroidal component in the
plasmoid morphology.

From a statistical point of view, plasmoids of smaller size, \ie
$4-6\,r_g$ along the filament and a diameter of $\sim 2-3\,r_g$, are
produced very frequently and over a timescale of $10-20\,M$. However, they
are either accreted by the black hole, or they are quickly destroyed by
the shear (see the blue clouds in Fig. \ref{fig:3Dplasm}). Much more
infrequently, \ie over a timescale of $\approx 500-700\, M$, plasmoids
with larger size, \ie $10-15\,r_g$, are generated and are thus able to
survive for longer times as clearly identifiable compact
features. Furthermore, also in the case of large plasmoids, their fate is
to be dissolved rather quickly. Indeed, in none of our simulations,
plasmoids are found to survive outside a spherical region of $\approx
30\,r_g$ from the event horizon.  Beyond this region, in fact, all
plasmoids have been converted to high-temperature clouds that
cool down with time.

\begin{figure}
	\begin{center}
		\includegraphics[width=0.48\textwidth]{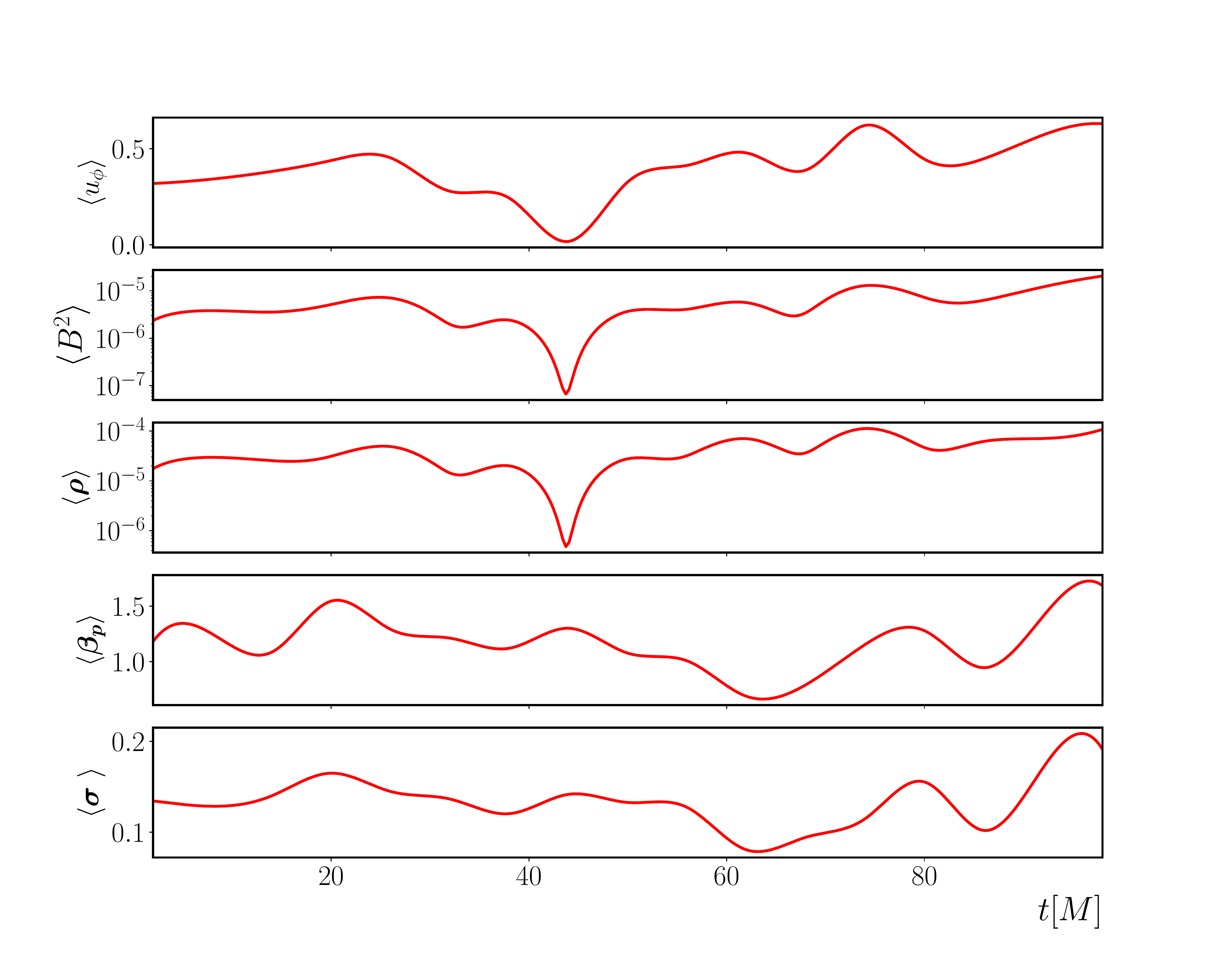}
	\end{center}
	\caption{Evolution of volume-averaged characteristic physical
	  quantities around the unbound plasmoid; shown from top to
	  bottom are the azimuthal velocity $u_{\phi}$, the
	  magnetic-field energy, the rest-mass density, the plasma
	  parameter $\beta_p$, and the magnetisation $\sigma$. Note that
	  the corresponding averaged quantities in the funnel are orders
	  of magnitude smaller; the data refers to model
	  \texttt{HB.id.0.5}.}
	\label{fig:evo}
\end{figure}

\subsection{Plasmoids: dynamics and flares}
\label{sec:plasmoids}

After having described where plasmoids are produced and their morphology
over time, it is now interesting to examine the possibility that these
plasmoids -- especially when they have sufficiently large energies and
orbit outwards around the black hole -- can be be associated with the
flares observed in Active Galactic Nuclei (AGN) \citep{Yuan2009,
  Giannios2013b, Younsi2015, Li2017flare}.

To this scope, the first important step is to locate and track the
trajectories followed by the plasmoids as they move outwards. This task,
however, is not trivial and far more demanding than in 2D simulations,
where the reduced dimensionality allows one first to find a plasmoid and
then track over time their evolution \citep[see, \eg, Fig. 9
  in][]{Nathanail2020}.

In a 3D simulation, on the other hand, plasmoids also have an azimuthal
motion and hence require a full 3D reconstruction of their motion. In
order to monitor the evolution of the plasmoids as a whole, we compute
averages over volumes containing the most energetic regions of the plasma
and compute there the evolution of the most relevant physical quantities
in the low-density funnel (see Fig. \ref{fig:evo}). Using model
\texttt{LB.id.0.0} as a reference, we monitor the evolution of the most
energetic plasmoids, \ie with $hu_t < -1.1$, and follow their evolution
for a timescale of $\approx 1000\,M$, which corresponds to a bit more
than $\approx 330\, {\rm min}$ when for a black hole with a mass as that
of Sgr~A*.

Figure \ref{fig:evo} reports the evolution of the most energetic
plasmoids, \ie with $h u_t > 1.01$, in terms of volume-averaged
quantities\footnote{For each scalar quantity $\Phi$, we define the
corresponding volume average over the unbound material as $\langle \Phi
\rangle:= \int_{hu_t > 1.0}\, \Phi\, dV / \int_{hu_t > 1.0}\, dV$. Note
that this is reasonable because the region with $hu_t > 1.0$ has a
compact support.} such as (from top to bottom): the azimuthal component
of the four-velocity $u_{\phi}/c$, the magnetic-field energy, the
rest-mass density, the plasma parameter $\beta_p:=p_{\rm g}/p_{\rm mag}$
(\ie the ratio between the fluid and magnetic pressures) and the
magnetisation around the plasmoid. Note that the azimuthal velocity
reaches values of $>0.5\ c$, which exceeds the local Keplerian velocity
from a radius of $\approx 3\,r_g$ and further out.

These energetic plasmoids can be very good candidates to explain the
observed orbital motions during flares from Sgr~A* \citep{Abuter2018b,
  Matsumoto2020}. Another important aspect of the super-Keplerian
plasmoids is that they get energised continuously, as seen from the
magnetic-field strength, the temperature and magnetisation evolution
(shown in the second, fourth and fifth row of Fig. \ref{fig:evo}
respectively). At the same time, the whole of the plasma in the funnel
slow down in the azimuthal direction, and tent to loose their
magnetisation, which however, was low from the beginning.

\section{Conclusions} 
\label{sec:con}

We have reported on three-dimensional GRMHD simulations of accretion
flows onto rotating black holes, either in the ideal-MHD limit or with a
finite physical resistivity. In all cases, the initial accretion flow is
produced by a magnetised MRI unstable torus which is seeded with poloidal
loops of alternating polarity. Using these simulations, we have shown
that magnetic reconnection in the vicinity of an astrophysical black hole
can efficiently produce macroscopic filamentary plasmoids with an
approximate length of $\sim 10-15\,r_g$ and a diameter of $\sim
3-5\,r_g$. They are generated approximately every $\approx 500-700\, M$,
and do not survive as compact structures outside a spherical region of
$\approx 30\,r_g$ from the event horizon. Plasmoids of smaller size, \ie
with a length of $4-6\,r_g$ and a diameter of $\sim 2-3\,r_g$, are
produced more frequently and over a timescale of $10-20\,M$. However,
they are either quickly destroyed by the shear or accreted by the black
hole.

Plasmoids can acquire large azimuthal velocities, exceeding the local
Keplerian velocity. Moreover, after their formation they can retain their
magnetisation or even increase it thanks to a continuous transfer of
energy through reconnection or the coalescence with other plasmoids. In
this way, they can be heated-up to relativistic temperatures,
$\Theta_e\approx 0.1-1$. During the evolution of the accretion flow,
multiple current sheets are formed inside the disc. However, due to the
low magnetisation, and the high density, this is not so efficient and
only locally insignificantly heats the plasma.

Interestingly, the overall evolution of the resistive model was found to
be close to the corresponding model evolved in the ideal-MHD limit, with
the exception of a lower variability in the accretion flow. Monitoring
the production of large-scale plasmoids, we could estimate indirectly a
reconnection rate of $\approx 0.01$ and measure the current density
reaching values $J/B>10^{2-3}$ in current sheets generated in the jet
sheath. In these regions, the local drift velocity was also found to be
$|(\vec{E}\times \vec{B})^r/B^2| \approx 10^{-2}$.

Flaring activity and intense variability has been observed in AGNs and
Sgr~A*, which hints to very rapid particle acceleration in a compact
region of a few gravitational radii \citep{Levinson2007, Begelman2008,
  Ghisellini2008, Giannios2009}. Magnetic reconnection in the vicinity of
a black hole can thus provide both requirements for an AGN flare. Our
simulations can set the base for such an analysis with underlying
realistic magnetic reconnection in general relativity and full 3D.
Furthermore, the generic behaviour of the release of magnetic energy in
the vicinity of a black hole can have implications to low accretion
radiatively inefficient flows such as Sgr~A*. The compact radio source at
the center of our galaxy has been observed with frequent flares
increasing more than two orders of magnitude from the usual flux
\citep{Ponti2017,Do2019}. Plasmoids produced in the models of our study
have shown a similar activity, and at times where macroscopic plasmoids
are formed, they acquire high azimuthal velocities which makes them
potential candidates to match and explain observations of orbiting hot
spots near the galactic center by the \citep{Abuter2018}.

Finally, we note that the analysis of how the plasmoids produced in
simulations of this type would lead to VLBI or infrared imaging requires
the coupling of our results to a complete imaging pipeline such as the
one employed by the EHT and GRAVITY collaborations. A detailed discussion
of such rendering will be presented in a future work

\section*{Acknowledgements}

AN and VM were supported by the Hellenic Foundation for Research and
Innovation (HFRI) under the 2nd Call for HFRI (Project Number:
00634). Support also comes from the ERC Advanced Grant ``JETSET:
Launching, propagation and emission of relativistic jets from binary
mergers and across mass scales'' (Grant No. 884631). The simulations
were performed on SuperMUC at LRZ in Garching, on the GOETHE-HLR cluster
at CSC in Frankfurt, and on the HPE Apollo Hawk at the High Performance
Computing Center Stuttgart (HLRS) under the grant numbers BBHDISKS and
BNSMIC.

\section*{Data Availability}
The data underlying this article will be shared on reasonable request to the corresponding author.

\section*{}
\bibliographystyle{mnras}
\bibliography{3Dplasmoid}


\label{lastpage}
\end{document}